# A Universal Deep Learning Force Field for Molecular Dynamic Simulation and Vibrational Spectra Prediction

**Shengjiao Ji[1], Yujin Zhang[2,\*], Zihan Zou[1], Bin Jiang[1,3], Jun Jiang[1,4], Yi Luo[4,5,\*] and Wei Hu[1,\*]**

[1] State Key Laboratory of Precision and Intelligent Chemistry, University of Science and Technology of China, Hefei, Anhui 230026, China

[2] International School for Optoelectronic Engineering, Qilu University of Technology (Shandong Academy of Sciences), Jinan, Shandong 250353, China

[3] School of Chemistry and Materials Science, University of Science and Technology of China, Hefei, Anhui 230026, China.

[4] Hefei National Research Center for Physical Sciences at the Microscale, University of Science and Technology of China, Hefei, Anhui 230026, China.

[5] Hefei National Laboratory, University of Science and Technology of China, Hefei, Anhui 230088, China.

* To whom correspondence should be addressed.

Email: _weihukth@gmail.com_ (W. Hu); _yiluo@ustc.edu.cn_ (Y. Luo); _zhangyujin@qlu.edu.cn_ (Y. Zhang)

## Abstract

Accurate and efficient simulation of infrared (IR) and Raman spectra is critical for molecular identification and structural analysis. Traditional quantum chemistry methods based on the harmonic approximation neglect anharmonicity and nuclear quantum effects, whereas ab initio molecular dynamics (AIMD) is prohibitively expensive. Here, we integrated our previously developed deep equivariant tensor attention network (DetaNet) with a velocity-Verlet integrator to enable fast and accurate machine learning molecular dynamics (MLMD) simulations for spectral prediction. Leveraging DetaNet's high-order tensor prediction capabilities, we first trained the model on the QMe14S dataset, which includes energies, forces, dipole moments, and polarizabilities for 186,102 small organic molecules, yielding a universal and transferable force field. We then simulated IR and Raman spectra using time-correlation functions derived from both MLMD and ring polymer molecular dynamics (RPMD) trajectories. Using isolated molecules, including polycyclic aromatic hydrocarbons, as a benchmark, we demonstrated that the DetaNet-based MD approach accurately captures anharmonic and nuclear quantum effects, producing spectra in excellent agreement with experimental data while achieving computational speedups of up to three orders of magnitude compared with AIMD. We further extended the framework to more complex systems, including molecular and inorganic crystals, molecular aggregates, and biological macromolecules such as polypeptides, with minimal fine-tuning. In all cases, DetaNet maintained high accuracy in simulating the IR and Raman spectra at substantially lower computational cost. Overall, this work presents a universal machine learning force field and a tensor-aware MLMD framework that enable fast and accurate dynamic simulations, as well as the subsequent prediction of IR and Raman spectra across diverse molecular and material systems.

# Main Text

## Introduction

Infrared (IR) and Raman spectroscopy techniques play crucial roles in molecular recognition and structural analysis owing to their distinctive vibrational fingerprints. In practice, accurate interpretation of experimental spectra often relies on theoretical simulations to enable reliable peak assignments and molecular characterization. Among the available approaches, the harmonic approximation based on quantum chemistry (QC) calculations is commonly used owing to its computational efficiency and straightforward interpretability[1-4]. Vibrational modes can be directly associated with specific molecular motions through frequency analysis. However, it neglects anharmonic effects, making it less accurate at describing high vibrational states, thermal fluctuations, and nonlinear couplings[5,6].

Alternatively, dynamical spectra can be obtained from molecular dynamics simulations by analyzing the time evolution of dipole moments or polarizabilities to calculate infrared or Raman spectra[7-9]. This approach naturally incorporates anharmonic effects, thermal fluctuations, and vibrational couplings, resulting in spectral positions, shapes, and intensities that are often closer to experimental observations. However, the accuracy of such spectra heavily relies on the underlying force field[10,11]. While ab initio molecular dynamics (AIMD) offers a more physically accurate description by performing simulations at the quantum chemical level, its high computational cost makes it impractical for large or complex molecular systems[12,13]. These limitations highlight the need for more efficient and accurate approaches to vibrational spectra simulation, especially for large-scale molecular recognition tasks.

Machine learning (ML) methods have become increasingly prominent in theoretical chemistry and vibrational spectral simulations, offering a way to bypass the high computational cost of traditional electronic structure calculations[14-18]. For example, Gastegger *et al.* developed a high-dimensional neural network potential (HDNNP)[19] trained on specific molecules, enabling machine learning molecular dynamics (MLMD) simulations and subsequent IR spectra predictions. They later proposed FieldSchNet[20], which incorporates external field interactions and solvation effects to improve dynamic IR simulations. Schütt *et al.* introduced the polarizable atom interaction neural network (PAINN)[21], which is designed to predict vectorial properties such as forces and dipole moments. By combining the scalar and vectorial outputs of the PAINN through cross-multiplication, they further enabled the prediction of polarizability tensors, allowing for the simulation of Raman spectra. However, these ML models are typically trained on individual molecules, limiting their transferability to novel chemical spaces. Moreover, they are not explicitly designed to directly predict high-order tensorial properties[22,23], such as Hessians or polarizability tensors, which are essential for accurate and generalizable Raman simulation.

In this work, we present a machine learning molecular dynamics protocol (DetaNet-MLMD) that integrates our previously developed deep equivariant tensor attention network (DetaNet)[24] with the velocity-Verlet algorithm to simulate dynamic infrared and Raman spectra. To incorporate nuclear quantum effects, we further extend the framework with RPMD. Evaluations across diverse systems, including isolated molecules, polycyclic aromatic hydrocarbons (PAHs), molecular and inorganic crystals, molecular aggregates, and polypeptides, demonstrate that DetaNet-MLMD and DetaNet-RPMD effectively address challenges in generalization and high-order tensorial property predictions. Together, they offer an efficient and transferable framework for dynamic simulations and real-time prediction of IR and Raman spectra.

## Results and Discussion

### Architecture of DetaNet-MD

We leveraged the high-order tensor prediction capabilities of DetaNet[24] and integrated it with the velocity-Verlet algorithm to establish a machine learning molecular dynamics (MLMD) protocol for simulating dynamic IR and Raman spectra. As illustrated in Figure 1a, the input systems ranged from isolated molecules and polypeptides without periodic boundary conditions to crystalline or aggregated structures with defined unit cells. Based on the input atomic coordinates with or without the parameters of the supercell, the initial momenta of each atom were assigned according to the Maxwell–Boltzmann distribution. Two integrators (Figure 1b) were subsequently used to iteratively update the atomic positions and velocities, thereby generating molecular dynamics trajectories using the SchNetPack[25] tools. During this process, DetaNet[24] was employed in each step to instantaneously compute molecular energies, atomic forces, dipole moments and polarizability (Figure 1c). All hyperparameter settings of DetaNet[24] are provided in Table S1 in Section 1 of the Supplementary Information.

For the MLMD simulations, the temperature was set to 300 K with a Nosé–Hoover chain[26] thermostat with a chain length of 3 and a relaxation time constant of 100 fs. Each MD simulation spanned 50 ps, with the initial 10 ps serving as an equilibration phase excluded from the analysis. The subsequent 40 ps of the MD trajectory were used for statistical analysis of structural, energetic, dipole, and polarizability properties. To balance the high-frequency mode descriptions and computational efficiency, we adopted system-specific time steps. Specifically, a 0.2-fs time step was used for all the isolated molecules, PAHs and ethanol aggregates, allowing precise sampling of fast vibrational dynamics. In contrast, a 0.5-fs time step was employed for 2-methylpyrazine aggregates, paracetamol and silicon dioxide crystals, and polypeptides to improve the computational efficiency[27]. The schematic structures and corresponding parameters of all the crystals are provided in Section 2 of the Supplementary Information. Both ethanol and 2-methylpyrazine aggregates consisted of 20 molecules, which were simulated in cubic boxes with side lengths of 12.469 Å and 14.478 Å, respectively. These crystal parameters were derived from the liquid density at room temperature.

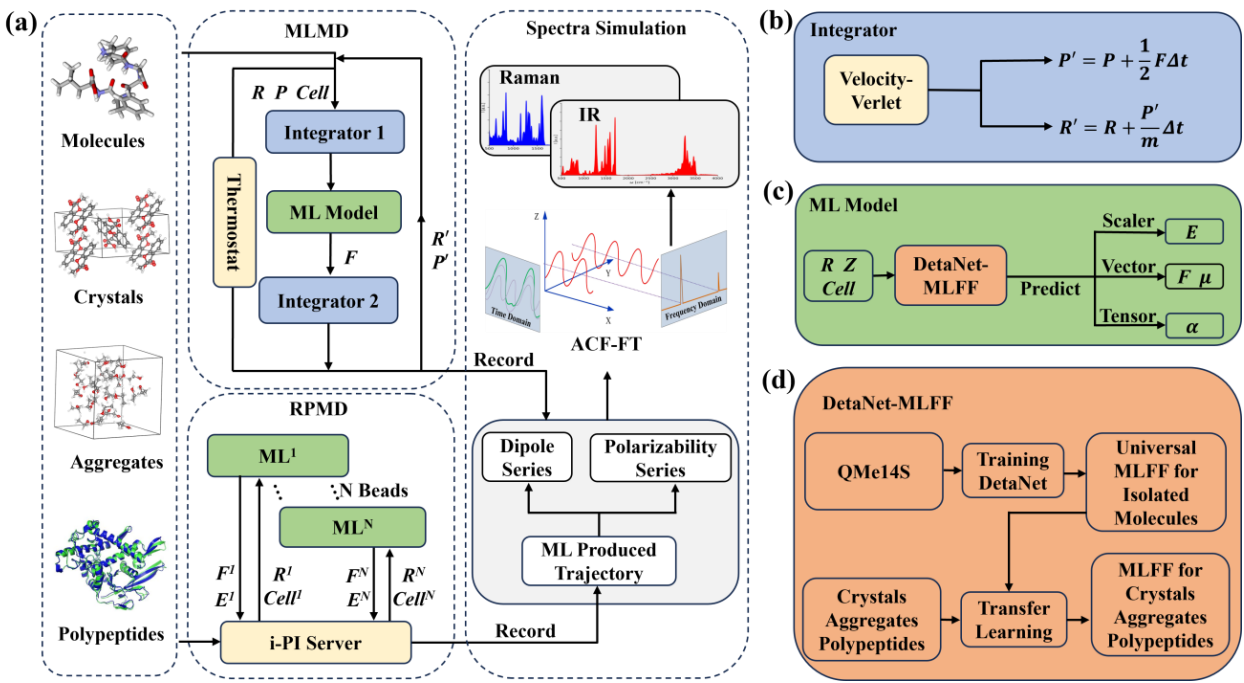

**Figure. 1| a** Architecture of DetaNet-based MLMD and RPMD simulations for isolated molecules, crystals, aggregates, and polypeptides, with a color-coded schematic highlighting individual components. **b** Schematic diagram of the velocity-Verlet integrator for updating the atomic position and momentum. **c** Schematic diagram of DetaNet[24] to predict the energy, force, dipole moment and polarizability. **d** Schematic diagram of training DetaNet[24] on the QMe14S[28] dataset to obtain a universal MLFF for isolated molecules and its extension via transfer learning to system-specific configurations for generating MLFFs applicable to crystals, aggregates and polypeptides.

To account for nuclear quantum effects (NQEs), we performed RPMD simulations using the SchNetPack[25] and i-PI[29] interfaces, with DetaNet[24] providing parallelized, real-time predictions of molecular energies and atomic forces across all the beads (replicas) in the ring polymer. The number of beads was set to 45 for isolated molecules and 48 for crystalline systems according to previous works[14,21]. Based on the DetaNet-predicted dipole moments and polarizabilities for each configuration along the MD trajectory, we performed Fourier transforms of the corresponding autocorrelation functions to convert the time-domain signal to a frequency-domain signal and obtained the anharmonic IR and Raman spectra, respectively.

**Isolated Molecules**

First, we trained DetaNet[24] on QMe14S[28] to obtain a universal force field (shown in Figure 1d). The QMe14S[28] dataset includes energy, force, dipole moment, and polarizability for 186,102 small isolated organic molecules, covering both equilibrium and nonequilibrium configurations sampled using atom-centered density matrix propagation (ADMP)[30] with the Gaussian 16 package[31]. We randomly split the QMe14S[28] dataset into training, validation, and test sets with percentages of 90%, 5%, and 5%, respectively. As shown in Figure 2a, DetaNet[24] accurately predicts atomic and molecular properties, achieving mean absolute errors of 0.0401 eV for energy, 0.0348 eV/Å for forces, 0.0253 D for dipole moments, and 0.2432 Å³ for polarizability, with all corresponding R² values exceeding 0.998.

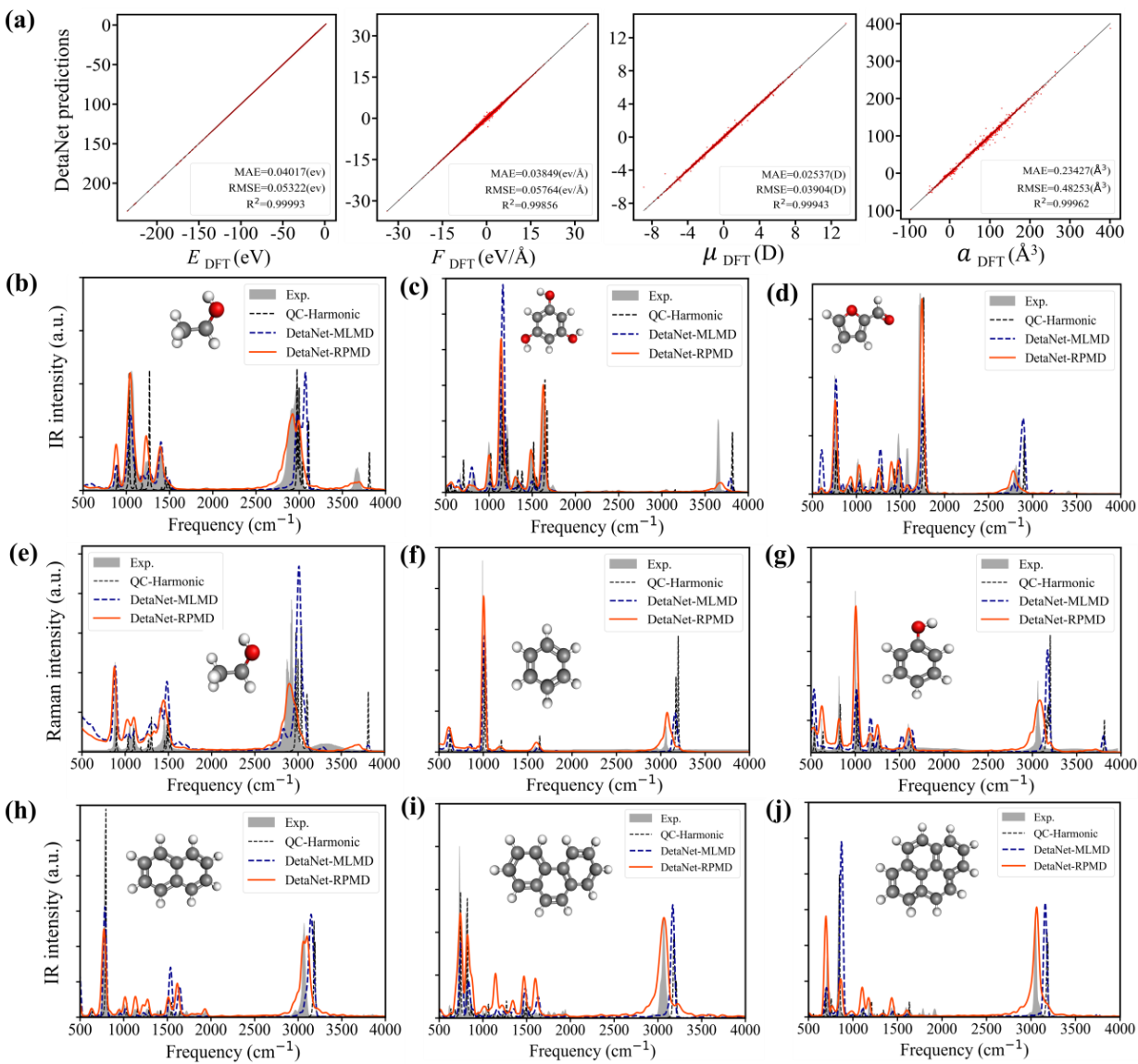

**Figure. 2| a** Error distributions and regression plots of QMe14S-trained DetaNet's predictions for the molecular energy, atomic force, molecular dipole moment and polarizability. **b–j** Comparison of vibrational spectra simulated using DetaNet-MLMD, DetaNet-RPMD, and QC-Harmonic methods, with experimental[32,33] spectra provided as reference: **b–d** IR spectra of ethanol, phloroglucinol, and furfural; **e–g** Raman spectra of ethanol, benzene, and phenol; **h–j** IR spectra of polycyclic aromatic hydrocarbons, including naphthalene, phenanthrene, and pyrene.

To evaluate the capability of our models to reproduce vibrational spectra, we selected five representative molecules from the test set and compared the simulated results from the quantum chemistry harmonic approximation (QC-Harmonic), DetaNet-MLMD, and DetaNet-RPMD with experimental IR and Raman spectra. Figures 2 b–g illustrates the IR spectra of ethanol, phloroglucinol, and furfural, as well as the Raman spectra of ethanol, benzene, and phenol. Notably, the QC-Harmonic approach yields only discrete vibrational frequencies and intensities, necessitating artificial broadening to generate continuous spectral line shapes. As such, it fails to capture anharmonic effects and often misrepresents the overall spectral envelope, typically introducing systematic blue shifts and overestimations in intensity. The typical deviations occur in the following regions:

1. IR region at 3600–3650 cm$^{-1}$: O–H stretching in ethanol and phloroglucinol (Figures 2b–c);

2. IR region at 2750–2850 cm$^{-1}$: Aldehyde C–H stretching in furfural (Figure 2d);

3. Raman region at 2850–2960 cm$^{-1}$: sp³ C–H stretching in ethanol (Figure 2e);

4. Raman region at 3000–3100 cm$^{-1}$: Aromatic C–H stretching in benzene and phenol (Figures 2f–g).

In contrast, DetaNet-MLMD accounts for anharmonic and thermal effects via molecular dynamics simulations, improving both peak positions and spectral profiles. As seen in Figures 2b–g, all C–H (2850–3100 cm$^{-1}$) and O–H (3600–3650 cm$^{-1}$) stretching modes are red-shifted relative to those of the QC-Harmonic mode, which aligns more closely with the experimental observations. However, some discrepancies persist. For example, DetaNet-MLMD overestimated the IR intensity of the aldehyde C–H stretch in furfural (Figure. 2d) and the Raman intensity of the aromatic C–H stretch in phenol (Figure. 2g).

DetaNet-RPMD further enhances spectral accuracy by incorporating NQE, resulting in improved peak positions, broadenings, and overall spectral agreement, particularly in high-frequency stretching regions. We quantitatively evaluated each method's performance by computing spectral similarities to experimental data using cosine, Pearson, and Spearman correlation coefficients (see Section 2 of the Supporting Information for formulas). As listed in Table 1, the models follow a consistent trend: DetaNet-RPMD > DetaNet-MLMD > QC-Harmonic for all the metrics. These results highlight the advantages of DetaNet-driven MLMD and RPMD frameworks in capturing complex vibrational dynamics and underscore their potential for high-throughput, accurate vibrational spectral predictions across diverse molecules.

Although DetaNet-RPMD shows improved agreement with the experimental IR and Raman spectra, some discrepancies remain. For instance, it slightly underestimated the IR intensities of the O–H stretching mode located at approximately 3600 cm$^{-1}$ for ethanol and phloroglucinol (Figures. 2b–c) but overestimated the Raman intensities of aromatic C–H stretching located at 3000–3100 cm$^{-1}$ in benzene (Figure. 2f). These deviations may arise from limitations in the exchange–correlation functional and basis set, which directly affect the quality of the potential energy surface and the force field used in dynamic simulations[34,35].

**Tab 1.** Spectral similarities (Cosine, Pearson, and Spearman correlation coefficients) between QC-Harmonic, DetaNet-MLMD, and DetaNet-RPMD simulations and experimental observations across various systems, including isolated molecules, PAHs, crystals, and aggregates. Detailed formulas are provided in Section 2 in the Supplementary Information.

| Different Systems | | Spectral Similarities | | | | | | | | |
|---|---|---|---|---|---|---|---|---|---|---|
| | | Cosine | | | Pearson | | | Spearman | | |
| | | QM | MLMD | RPMD | QM | MLMD | RPMD | QM | MLMD | RPMD |
| Isolated Molecules | Ethanol (IR) | 0.45 | 0.66 | 0.94 | 0.38 | 0.60 | 0.92 | 0.69 | 0.70 | 0.80 |
| | Phloroglucinol (IR) | 0.45 | 0.62 | 0.90 | 0.38 | 0.59 | 0.89 | 0.63 | 0.73 | 0.85 |
| | Furfural (IR) | 0.14 | 0.62 | 0.63 | 0.08 | 0.58 | 0.65 | 0.76 | 0.79 | 0.85 |
| | Naphthalene (IR) | 0.32 | 0.56 | 0.79 | 0.29 | 0.51 | 0.76 | 0.55 | 0.57 | 0.74 |
| | Phenanthrene (IR) | 0.28 | 0.29 | 0.76 | 0.22 | 0.23 | 0.73 | 0.52 | 0.56 | 0.57 |
| | Pyrene (IR) | 0.33 | 0.34 | 0.66 | 0.52 | 0.53 | 0.61 | 0.49 | 0.50 | 0.51 |
| | Ethanol (Raman) | 0.23 | 0.46 | 0.60 | 0.14 | 0.39 | 0.67 | 0.37 | 0.48 | 0.52 |
| | Benzene (Raman) | 0.20 | 0.89 | 0.90 | 0.15 | 0.81 | 0.89 | 0.42 | 0.50 | 0.55 |
| | Phenol (Raman) | 0.26 | 0.59 | 0.70 | 0.13 | 0.44 | 0.61 | 0.41 | 0.51 | 0.52 |
| Molecular Crystals | Paracetamol (IR) | | 0.70 | 0.87 | | 0.54 | 0.77 | | 0.80 | 0.85 |
| | Paracetamol (Raman) | | 0.70 | 0.77 | | 0.64 | 0.76 | | 0.71 | 0.77 |
| Molecular Aggregates | Ethanol (IR) | | 0.89 | 0.90 | | 0.87 | 0.88 | | 0.85 | 0.86 |
| | 2-Methylpyrazine (IR) | | 0.73 | 0.83 | | 0.67 | 0.79 | | 0.68 | 0.73 |
| Polypeptides | LeuEnk (IR) | | 0.80 | | | 0.66 | | | 0.76 | |

To investigate the impact of functionals and basis sets on spectral accuracy, we calculated the atomic forces and dipole moments for 2,000 molecular dynamics (MD) configurations of ethanol and phloroglucinol using various combinations of functionals (BP86[36], PBE[37], and M06-2X[38]) and basis sets (6-31G**[39] and 6-311++G**[40]). These data were subsequently used to fine-tune the QMe14S-pretrained DetaNet[24] model through transfer learning, followed by AIMD simulations. As shown in Figure S1 (Section 4, Supplementary Information), the simulated spectra closely reproduced the experimental results, particularly for vibrational modes below 2000 cm$^{-1}$. However, the accuracy varied considerably depending on the choice of functional and basis set. Notably, the M06-2X[38] functional group, which includes long-range interactions, outperformed PBE[37] and BP86[36] in modeling the O–H stretching modes of ethanol and phloroglucinol. These results underscore the strong dependence of dynamic vibrational spectra on the quality of the underlying potential energy surface and the precision of trajectory sampling. As such, dynamic IR and Raman spectra simulations serve not only as probing tools but also as valuable benchmarks for assessing the reliability of force fields.

To evaluate the generalizability of DetaNet-based models to unseen chemical systems, we evaluated their ability to reproduce IR spectra for a set of PAHs, which are highly important for astronomy, environmental science, materials science, and biological health. As shown in Figures 2h–j, both DetaNet-MLMD and DetaNet-RPMD produced IR spectra of naphthalene, phenanthrene, and pyrene molecules that closely match experimental measurements, significantly outperforming the QC-Harmonic method. This improvement is particularly evident in the high-frequency C–H stretching region ($\sim$3000–3200 cm$^{-1}$), where QC-Harmonic systematically overestimated vibrational frequencies. In contrast, both DetaNet[24] models accurately predicted peak positions and better reproduced the overall spectral shape.

**Molecular Crystals**

We assessed the transferability of the DetaNet[24] model to diverse and complex systems, including organic and inorganic crystals, molecular aggregates, and polypeptides. Taking the paracetamol crystal as an example, we first carried out ab initio molecular dynamics (AIMD) simulations using the CP2K[41] software package at the PBE0 [42]/pob-TZVP[43] level, generating 35,000 trajectory frames. As shown in Figure 3a, direct training (hereafter referred to as de Novo Learning, DNL) on these 35,000 configurations yielded nearly perfect agreement with the DFT reference forces (MAE $\approx$ 0.00798 eV Å$^{-1}$, R$^2$ $\approx$ 0.99987). Given the high computational cost of AIMD, we explored reducing the dataset size to minimize the expense of subsequent spectral simulations. Training the DNL with only 2,000 configurations (Figure 3b) led to noticeably larger prediction errors and clear deviations from the DFT data (MAE $\approx$ 0.04946 eV Å$^{-1}$, R$^2$ $\approx$ 0.98739). Remarkably, transfer learning (TL) from a QMe14S[28] pretrained model to the same 2,000 configurations (Figure 3c) achieved an accuracy comparable to that of the large-scale DNL model (MAE $\approx$ 0.01961 eV Å$^{-1}$, R$^2$ $\approx$ 0.99912), demonstrating that TL can drastically reduce data requirements while maintaining high fidelity.

In addition to improving prediction accuracy, TL significantly accelerates model convergence[44-47]. As shown in Figure 3d, TL on 2,000 configurations achieved substantially lower MAEs and higher R$^2$ values within the first few hundred epochs, whereas DNL required far more iterations to reach similar accuracy. Comparable results were obtained for silicon dioxide crystals (shown in Figures 3e–h), highlighting the pronounced benefits of leveraging prior knowledge from a pretrained model.

To systematically assess the impact of dataset size on the TL performance, we fine-tuned the QMe14S[28] model with 500, 1,000, 2,000, 5,000, 10,000, 20,000, and 35,000 configurations (Figure 3i) and tested it on a fixed set containing 1,750 configurations. The model fine-tuned with 2,000 configurations achieved a validation set MAE of 0.01961 eV Å$^{-1}$ and an R$^2$ of 0.99912, representing an excellent balance between accuracy and computational cost. These results indicate that for systems similar to the pretraining set, a TL with approximately 2,000 configurations is sufficient to achieve near-saturated performance, providing an efficient strategy for force field adaptation.

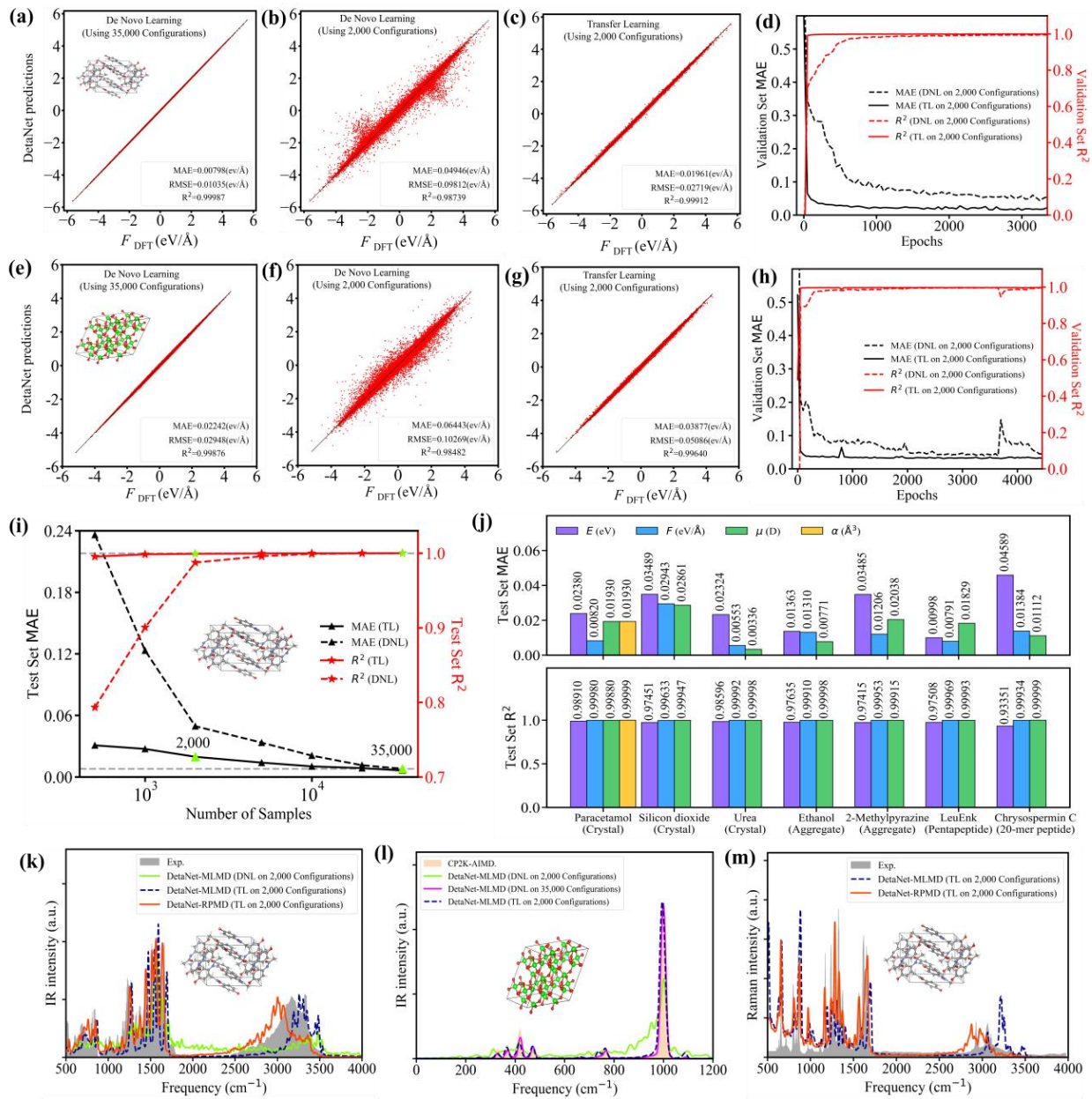

**Figure. 3| a–b** Mean absolute errors (MAEs) and coefficients of determination (R²) for atomic forces predicted by de novo learning (DNL) using 35,000 and 2,000 AIMD configurations of paracetamol crystals. **c** MAE and R² for atomic forces, obtained through transfer learning (TL) from the QMe14S[28] pretrained model and fine-tuned on 2,000 configurations. **d** Comparison of convergence behavior between DNL and TL using 2,000 configurations. **e–h** Analogous results for silicon dioxide crystal (cf. panels **a–d**). **i** Effect of fine-tuning dataset size (500–35,000 configurations) on TL performance. **j** MAEs and R² values for energy, forces, dipole moments, and polarizabilities predicted by the TL model across six representative systems: paracetamol crystal, urea crystal, silicon dioxide crystal, ethanol aggregates, 2-methylpyrazine aggregates, the LeuEnk pentapeptide, and a 20-mer peptide. DFT results serve as the reference. **k–m** Comparison of vibrational spectra simulated using

DetaNet-MLMD, DetaNet-RPMD, and QC-Harmonic methods, with experimental[33,48] spectra provided as reference: (**k**) IR spectra of the paracetamol crystal; (**l**) IR spectra of the silicon dioxide crystal; (**m**) Raman spectra of the paracetamol crystal.

We further extended TL to predict the energy, dipole moment, and polarizability for the paracetamol crystal, silicon dioxide crystal, urea crystal, ethanol aggregates, 2-methylpyrazine aggregates, the LeuEnk (pentapeptide), and a larger 20-mer polypeptide. All AIMD reference simulations were performed using CP2K, with the B3LYP[49]/pob-TZVP[43] functional for organic crystals and aggregate systems and PBE[37]/DZVP-MOLOPT-SR-GTH[50] for the silicon dioxide crystal and peptides, following prior literature protocols[51]. As shown in Figure 3j, the resulting TL-based force fields achieved outstanding accuracy, with R² values exceeding 0.996 across all systems for force predictions and similarly high performance for dipole moment and polarizability predictions.

The DetaNet-MD-simulated IR spectra of paracetamol and silicon dioxide crystals, as well as the Raman spectra of paracetamol, are presented in Figures 3k–m. As illustrated in Figures 3k and 3l, the DNL-trained force field using 2,000 configurations (green line) failed to accurately reproduce the IR spectra for both organic and inorganic systems. In contrast, the TL model fine-tuned on only 2,000 configurations achieved spectral accuracy comparable to that of the DNL model trained on 35,000 configurations (see Figure S4 in the Supporting Information for details). Accordingly, the TL-trained force field is employed throughout the following discussions to evaluate the performance of the DetaNet-MD models.

As shown in Figures 3k–m, both DetaNet-MLMD and DetaNet-RPMD showed reasonable agreement with the experimental results, but DetaNet-RPMD offered clear improvements in regions where NQEs are significant. For instance, DetaNet-MLMD produced blue-shifted peaks for the coupled benzene stretching and N–H bending mode (1670 cm⁻¹) in paracetamol (as shown in Figures 3k and 3m). DetaNet-RPMD corrected these shifts and yielded results closer to the experimental values. In the high-frequency region, DetaNet-MLMD overestimated the peaks near 3200 and 3330 cm⁻¹, which correspond to the N–H and O–H stretching modes in paracetamol. Although DetaNet-RPMD introduced a slight red shift compared with the experiment, it still offered better alignment overall. As listed in Table 1, DetaNet-RPMD consistently yielded higher spectral similarity coefficients than DetaNet-MLMD for paracetamol, confirming its superior ability to capture all the vibrational features.

**Molecular Aggregates**

Taking the 2-methylpyrazine aggregates and ethanol aggregates as representative nonperiodic systems, we evaluated the performance of DetaNet-MLMD and DetaNet-RPMD in capturing both structural and spectroscopic properties for aggregated but nonperiodic systems. To assess structural accuracy, we first calculated the radial distribution functions (RDFs) and mean square displacements (MSDs) for both systems. As shown in Figures 4a–b, the N–H and O–H RDFs predicted by DetaNet-MLMD closely match those obtained from CP2K-AIMD simulations, indicating that the model accurately reproduces local hydrogen-bonding structures. Moreover, the MSDs of 2-methylpyrazine and ethanol aggregates exhibited nearly identical trends to the AIMD results, and the computed diffusion coefficients are statistically indistinguishable (Figures 4c–d), confirming DetaNet's fidelity in dynamic simulations.

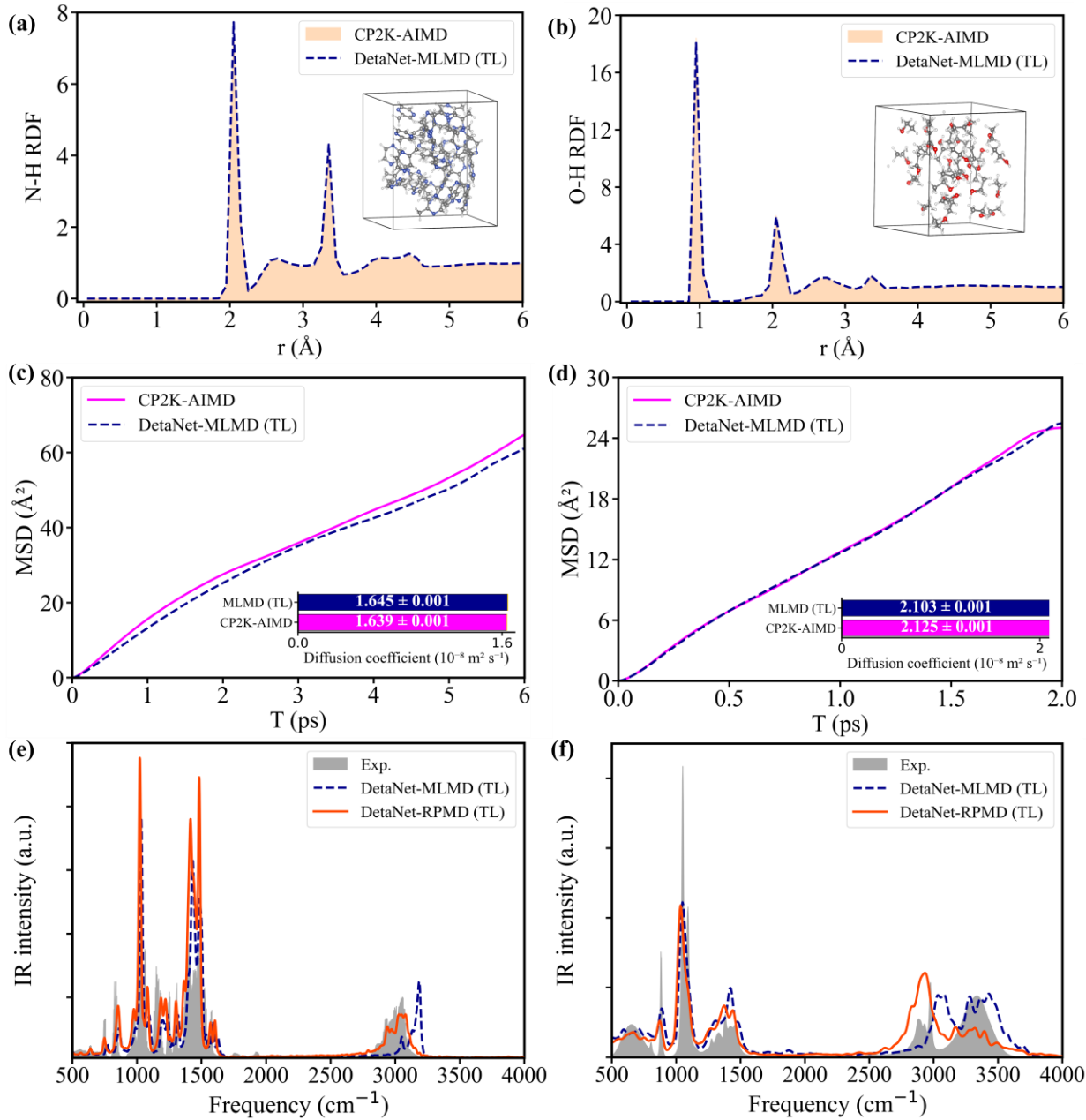

**Figure. 4| a–b** Radial distribution functions (RDFs) for N–H and O–H interactions in 2-methylpyrazine aggregates and ethanol aggregates, computed using DetaNet-MLMD and compared with CP2K-AIMD results. **c–d** Mean square displacements (MSDs) of the same systems predicted by DetaNet-MLMD and CP2K-AIMD.

**e–f** IR spectra of 2-methylpyrazine aggregates and ethanol aggregates simulated using DetaNet-MLMD and DetaNet-RPMD, with experimental[32,48] spectra provided for comparison.

For the vibrational spectra, both DetaNet-MLMD and DetaNet-RPMD reproduced the experimental peak positions and intensities with reasonable accuracy (shown in Figures 4e–f). However, DetaNet-RPMD yielded consistently better agreement, especially in the high-frequency region. For example, DetaNet-MLMD slightly overestimated the frequencies and produced narrower peaks near 3100 cm$^{-1}$, which correspond to the C–H stretching modes in 2-methylpyrazine aggregates. DetaNet-MLMD also overestimated the frequencies of the O–H stretching band between 2900 cm$^{-1}$ in ethanol aggregates. In contrast, DetaNet-RPMD accounted for NQEs, resulting in red-shifted and broader peaks that more closely match the experimental IR spectra. As listed in Table 1, the cosine, Pearson, and Spearman spectral similarity coefficients obtained from DetaNet-RPMD were consistently higher than those from DetaNet-MLMD, demonstrating the superior accuracy of DetaNet-RPMD in reproducing experimental spectra.

**Polypeptides**

We evaluated the accuracy and transferability of the DetaNet-MLMD model for biomolecular vibrational spectroscopy. Notably, owing to the substantial computational cost, DetaNet-RPMD is not feasible for large biological systems. We first examined the effect of the cutoff radius on capturing long-range polypeptide interactions. As shown in Figures 5a–c, the cutoff radius significantly influenced the accuracy of the machine learning force fields. Increasing the cutoff from 5 Å to 8 Å substantially improved the correlation between the predicted and DFT reference forces, reduced the prediction error, and underscored the importance of accounting for longer-range interactions[52-55]. Furthermore, the IR spectrum of a representative pentapeptide computed using DetaNet-MLMD is in excellent agreement with the experimental data, accurately reproducing both band positions and relative intensities. Notably, the simulated intensity ordering, amide II (1500–1600 cm$^{-1}$) > amide I (1600–1700 cm$^{-1}$) > amide III (1200–1350 cm$^{-1}$)[56], is consistent with experimental observations. A quantitative comparison using spectral similarity metrics further confirms the model's accuracy, as listed in Table 1.

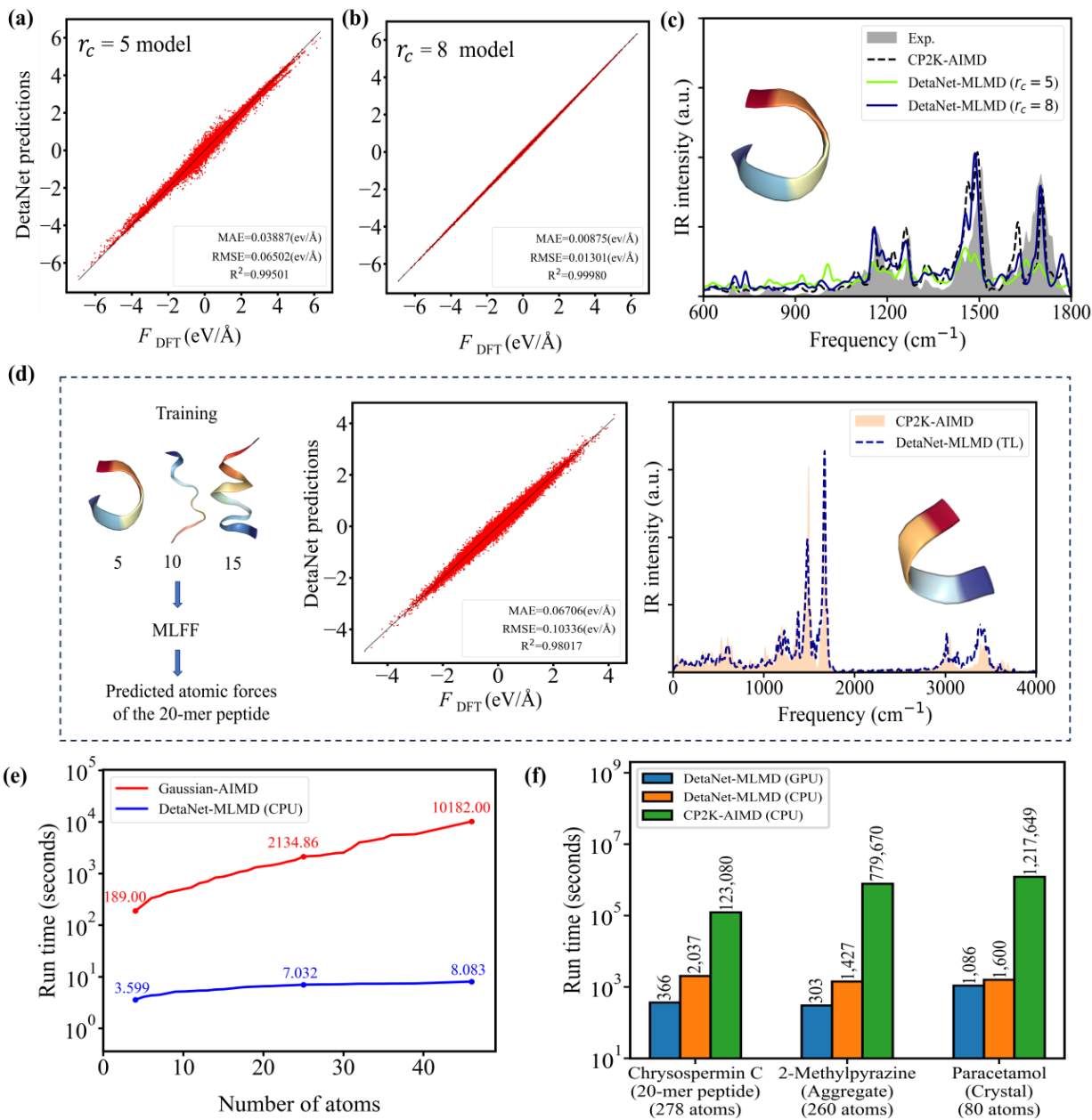

**Figure. 5**| **a–b** Mean absolute errors (MAEs) and coefficients of determination (R²) for pentapeptide atomic forces predicted by transfer learning (TL) using $r_c = 5$ and $r_c = 8$ models. **c** Infrared (IR) spectra of a representative pentapeptide simulated using DetaNet-MLMD, CP2K-AIMD, $r_c = 5$ and $r_c = 8$, with experimental[57,58] data shown for reference. **d** Schematic illustration of DetaNet[24] training workflow on peptides of varying lengths (5-, 10-, and 15-mer), followed by the evaluation of its predictive performance on an unseen 20-mer peptide. The results include both the machine learning force field (MLFF) predictions and the corresponding simulated IR spectra. **e–f** Benchmarking the computational efficiency of DetaNet-MLMD compared with that of Gaussian-AIMD for isolated molecules and that of CP2K-AIMD for molecular crystals, aggregates, and polypeptides in dynamic vibrational spectroscopy simulations.

To further evaluate the transferability of the DetaNet[24] force field, we extended our analysis to larger helical 20-mer peptides that contain 278 atoms (inset, Figure 5b), despite the absence of corresponding experimental spectra. Here, we avoided any structural data from the 20-mer peptide during training. We pretrained DetaNet[24] on the QMe14S[28] dataset and performed fine-tuning using shorter peptide fragments, including 5-mer, 10-mer, and 15-mer structures (illustrated in Figure 5d). Despite this limited training scope, the resulting model demonstrates excellent predictive accuracy on the unseen 20-mer system, achieving a force MAE of 0.067 eV/Å and an $R^2$ of 0.98 relative to the DFT-calculated forces (Figure 5d). Based on this transferable force field, we computed the IR spectrum of the 20-mer peptide. As shown in Figure 5d, the predicted peak positions are nearly identical to those from a DetaNet[24] model fine-tuned on 2000 structures of the 20-mer itself, further validating the generalizability of the model. However, the predicted amide I intensity (1600–1700 cm$^{-1}$) exceeded the amide II intensity (1500–1600 cm$^{-1}$), deviating from the CP2K-AIMD result. This discrepancy suggests that while the force field exhibits strong transferability from small to large biomolecular systems, the learned dipole moment surface may require additional refinement to ensure consistent spectral intensities across scales.

**Computational Efficiency**

To evaluate the computational efficiency of the DetaNet-MLMD framework, we benchmarked its runtime performance against that of conventional AIMD methods, using Gaussian for isolated molecules and CP2K for the other extended systems. As shown in Figure 5e, DetaNet-MLMD achieved over 500-fold speedup compared to Gaussian-AIMD across a range of isolated molecular systems. This performance advantage became increasingly pronounced with increasing molecular size. For example, while Gaussian-AIMD required more than 10,000 seconds to simulate a 45-atom molecule, DetaNet-MLMD completed the same task in just 8.08 seconds.

For more complex systems, such as molecular crystals (paracetamol, 80 atoms), molecular aggregates (2-methylpyrazine, 260 atoms), and large biomolecular assemblies (chrysospermin C, 278 atoms), DetaNet-MLMD delivered speedups ranging from 750 to over 1000 relative to CP2K-AIMD (shown in Figure 5f). For example, generating a 2-ps trajectory with dipole and polarizability calculations for the paracetamol crystal required only 1,600 seconds with DetaNet-MLMD, compared to over 1.2 million seconds (~two weeks) using CP2K on the same CPU architecture.

Finally, we evaluated the scalability of DetaNet[24] through MD simulations of expanded polypeptide systems. Benchmark tests demonstrate the framework's capability to simulate proteins containing 1,489 peptides (9,244 atoms) on a standard 256 GB memory CPU node while maintaining DFT-level accuracy in computed properties, including Hessian matrices, polarizability tensors, and resulting IR/Raman spectra. This computational efficiency, combined with quantum-mechanical fidelity, positions DetaNet-MD as a promising tool for large-scale biomolecular simulations.

## Conclusion

Overall, we proposed DetaNet-MLMD and DetaNet-RPMD to establish a machine learning molecular dynamics protocol for simulating dynamic IR and Raman spectra. To enhance its generalizability, we trained DetaNet[24] on our QMe14S[28] dataset, which includes energy, force, dipole moment, and polarizability data for approximately 186,102 molecules in both equilibrium and nonequilibrium geometries. By integrating DetaNet[24] with the velocity-Verlet algorithm, we enabled MLMD simulations and further incorporated RPMD to account for nuclear quantum effects. Evaluations across various systems show that DetaNet-MLMD generalizes well to isolated molecules, delivering highly accurate dynamic spectra even for species beyond the training set. For more complex systems such as molecular aggregates, crystals, and polypeptides, we found that applying transfer learning from the QMe14S-pretrained DetaNet[24] to a small set of 2,000 system-specific configurations significantly improved the accuracy of the machine learning force field compared to training the model from scratch. By overcoming the generalization limitations and tensorial prediction challenges, DetaNet-MLMD and DetaNet-RPMD offer an efficient and transferable framework for real-time simulation of dynamic IR and Raman spectra across a wide range of molecular systems, supporting advanced spectral analysis and molecular structure recognition.

## Data Availability

All datasets used in this document are publicly available. Source data for Figures 2–5 are available with this manuscript.

The original QMe14S[28] dataset is available from https://figshare.com/s/889262a4e999b5c9a5b3.

The datasets used for training crystals, aggregates, and polypeptides properties, as well as the molecular trajectories and spectral data generated from DetaNet-MLMD and DetaNet-RPMD simulations, are available at https://figshare.com/s/043b1ace6546b4221a43.

## Code Availability

The DetaNet-MD package is available from https://github.com/WeiHuQLU/DetaNet-MD

## Acknowledgments

We acknowledge the grant support from the National Natural Science Foundation of China 22573055 (W.H.), the Natural Science Foundation of Shandong Province ZR2023MA089 (Y.Z.), the Program for Introduced Innovation Teams from the New Collegiate 20 Items of Jinan 202228031 (Y.Z.), and the Science and Technology Support Program for Youth Innovation in the Universities of Shandong Province

## Author Contributions

W.H. and S.J. conceived the research, designed the DetaNet-MLMD and DetaNet-RPMD models and performed all data analyses. W.H., Y.L., and Y.Z. jointly supervised the work from the model design to the data analysis. W.H., S.J., Y.Z., Z.Z., B.J., J.J., and Y.L. interpreted the data. All authors contributed to the writing of the manuscript.

## Competing interests

The authors declare no competing interests.